\documentclass[twocolumn, amsmath, amssymb]{revtex4-1}

\usepackage{graphicx}
\usepackage{amsmath,amssymb}

\newcommand{\Z}{\mathbb{Z}}

\newcommand{\lseq}[2]{{\{#1\}}_{#2}}
\newcommand{\ket}[1]{\mathinner{|{#1}\rangle}}
\renewcommand{\frac}[2]{#1/#2}

\begin{document}
\title{Publicly Verifiable Blind Quantum Computation}
\author{Kentaro Honda*}
\affiliation{
School of Information Science and Technology, the University of Tokyo, 7-3-1 Hongo, Bunkyo-ku, Tokyo 113-8656, Japan}
\email[Email:]{honda@is.s.u-tokyo.ac.jp}

\begin{abstract}
Blind quantum computation protocols allow a user with limited quantum technology to delegate an intractable computation to a quantum server while keeping the computation perfectly secret.
Whereas in some protocols a user can verify that calculated outcomes are correct, a third party cannot do this, which allows a dishonest user or owner to benefit illegally.
I propose a new blind quantum computation protocol with a new property called public verifiability, which enables any third party to assure that a party does not benefit from attempted deception.
\end{abstract}

\maketitle

Blind quantum computation (BQC) protocols~\cite{BFK09,BFKW13,BKBFZW12,DKL12,FK13,GMMR13,GKW15,HM15,M13,M14,MF12,MF13,MPF13} allow Alice, who has no quantum computer, to use quantum computation without buying one or sacrificing her privacy.
In these protocols, if Alice has weak quantum devices, such as a single qubit generator, she will be able to delegate an intractable computation to Bob's quantum server, keeping the computation perfectly secret.
Several BQC protocols~\cite{BFK09, BFKW13, FK13, GMMR13, GKW15,HM15,M13, M14} have been given an additional property, called verifiability.
Although he cannot obtain the outcome of the computation, evil Bob will try to deceive Alice and send an incorrect outcome.
The verifiability property enables her to detect whether a given outcome is correct, even though she cannot recheck the outcome itself.
Existing verifiable BQC protocols were given the property in essentially the same method as follows~\cite{FK13}.
For traps, Alice secretly adds independent trivial parts to her desired computation.
She knows the expected outcomes of those computations.
Because Bob does not know where the traps are placed, there is only a small probability that he can tamper with her computation without being caught in one.

When Alice detects the incorrect outcome, it is natural that she rebuke Bob for cheating and reject paying his server fee.
However, if the rejection is allowed, a further problem may arise.
Assume the rejection of the fee is allowed, and Bob is honest, but Alice wants to be a free rider (i.e., she wants to obtain the outcome of the computation without paying for it).
Her winning strategy is simple: follow the verifiable BQC protocol and say ``Bob cheats'', ignoring the outcomes of the traps.
Unfortunately, because the aforementioned verification method uses her private information, existing verifiable BQC protocols provide methods of detecting the incorrect outcome only for Alice; thus, a third party, even a court, cannot verify whether Bob really cheats and cannot settle the dispute.
Even if a court orders her to give evidence against Bob when the conflict arises, BQC protocols allow the would-be free rider to forge it, because the protocols perfectly conceal what she does.
Indeed, a no-go theorem suggests that it is impossible to unconditionally resolve the conflict~\cite{M97}.
However, it remains possible to resolve it computationally.
Here, with the aid of classical cryptography, I propose a new verifiable BQC protocol based on an existing verifiable BQC protocol.
Without any additional assumption, this protocol preserves the properties of the original protocol (e.g., perfect security, unconditional verifiability, and the probability that Alice will detect Bob's attempt to deceive).
Furthermore, if Alice has no quantum memory, the protocol satisfies a further property that any third party with a classical computer can unconditionally detect Bob's attempt to deceive and computationally detect Alice's deception.
I call this property ``public verifiability''.
Because of this property, any third party can judge whether Alice should pay.
My protocol is the first verifiable BQC protocol with such a property and enables users in an era of first-generation quantum computers to use them without absurd costs.

Before revealing my protocol, I briefly review the verifiable BQC Fitzsimons--Kashefi~(FK) protocol~\cite{FK13} on which my protocol is based.
That protocol uses measurement-based quantum computation~(MBQC)~\cite{RB01} and proceeds as follows. 
First, Alice sends Bob single qubit states, from which Bob constructs a graph state.
Almost all single qubit states are rotated by random angles, but she secretly chooses the locations of trap qubits, and all qubits that neighbor any of the traps are set to $\ket{0}$ or $\ket{1}$ so that the trap qubits are separated from the other qubits.
Then, for each qubit, Alice decides a measurement angle and sends it to Bob, who measures the qubit in the angle and sends back the measurement result.
The measurement angle $\delta_i$ is adjusted to not only the previous measurement results $\lseq{b_j}{j < i}$ as the usual feed-forward process in MBQC, but also to the random rotation angle $\theta_i$ by
\begin{equation}
  \delta_i = {(-1)}^{b_{X_i}+r_{X_i}}\phi_i + {\theta_i}' + \pi\sum_{j \in Z_i} (b_j+r_j) \pmod{2\pi} \label{feedforwardeq}
\end{equation}
where $\phi_i$ is the original computational angle, $r_j$ is a random bit hiding the measurement result, ${\theta_i}' = \theta_i + \pi r_i$, and subscripts $X_i, Z_i$ are sets determined by the graph and denote bit summations.
Alice sets $\phi_t = 0$ for each trap qubit.
Finally, she accepts the outcome of the computation if $b_t = r_t$ for any trap $t$.
Notice that the FK protocol requires Alice to have no quantum device except for a single qubit generator.

BQC protocols are required to be blind (if Alice respects the protocol, Bob cannot learn anything about her computation no matter what he does) and correct (if all parties respect the protocol, it does not abort and Alice obtains the correct outcome).
In the FK protocol, the random angles $\lseq{\theta_i}{}$ and random bits $\lseq{r_i}{}$ make the protocol blind, and the angle adjustment~(\ref{feedforwardeq}) makes it correct.
Note that $\delta_{t}$ is equal to $\theta_{t} + \pi r_{t}$ and thus $b_{t}$ should equal $r_{t}$ for any trap $t$.
Moreover, the secret separations of the trap qubits from the others provide unconditional verifiability.
A BQC protocol is considered $\epsilon$-verifiable if the probability of overlooking Bob's attempt to deceive does not exceed $\epsilon$ regardless of what he does.
Although the FK protocol does not specify which graph state is used, with the dotted-complete graph state and a computation encoded in a fault-tolerant manner, the FK protocol achieves $\epsilon = {(\frac{2}{3})}^{\lceil\frac{2}{5}d\rceil}$ with security parameter $d$~\cite{FK13}.

The purpose of this paper is to determine a novel BQC protocol satisfying a new property: public verifiability.
Let us first describe its setting.
In addition to Alice and Bob, I consider a third party who can collect all classical messages sent between Alice and Bob.
Based on the messages alone, the third party judges whether the correct outcome is obtained by Alice.
When Alice and Bob broadcast all their classical messages, anyone can work as a third party, and I therefore do not consider a cheating third party.
Alice or Bob may be evil, but I ignore the case where both are evil.
Evil Alice attempts to make a third party judge that the outcome is incorrect and to learn something about the delegated computation.
She can secretly embed her desired computation within a larger computation, and this may be enough for her to obtain some partial information.
Moreover, she may know some information about the computation before starting the protocol, and she can use this to learn the desired information.
I assume that her computational ability is the same as that of a classical computer, and hence the outcome of the delegated computation is classical.
This assumption is justified by the fact that she delegates her computation via a BQC protocol.
If she is powerful, she can compute it herself.
Public verifiability claims that an evil party will fail to achieve its purpose.
Let $\xi$ be a number less than one.
A BQC protocol is said to be $\xi$-publicly verifiable if (1) the probability that a third party wrongly judges that Alice obtains the correct outcome is bounded by $\xi$ and (2) when Bob is honest and a third party judges that Alice does not obtain the correct outcome, Alice cannot obtain any non-negligible partial information about the delegated computation.
The precise definition, with a cryptographic flavor, can be found in the Appendix.
One may think that although Alice has no quantum computer, she will still be able to use another quantum server to attack the protocol and obtain the outcome.
However, in that case, she cannot benefit from the deception.
If the server requires her to use a publicly verifiable BQC protocol, she has to pay the server for the outcome, as she cannot obtain it otherwise.
Therefore, I ignore this type of attack.

I will now describe my protocol, which uses a classical probabilistic public-key encryption scheme.
The protocol does not depend on which encryption scheme is used, provided that it satisfies four requirements, which I will impose later.
For simplicity, I assume that when receiving messages, Alice or Bob checks the validity of messages and aborts if any of them is found to be invalid.
The protocol runs as follows. 
(I) Alice selects a graph $G$ and randomly sets the locations of traps $T$ from the vertices $V$.
She also chooses uniformly randomly $\lseq{d_i}{i \in N_{G}(T) }$ and $\lseq{\theta_i}{i \in V}$ from $\{0, 1\}$ and $\{\frac{k\pi}{4} \mid k = 0, \ldots, 7\}$, respectively, where $N_{G}(T)$ is the neighborhood of $T$.
She announces the graph $G$ and sends Bob single qubit states $\lseq{\ket{q_i}}{i \in V}$ where
\begin{equation}
  \ket{q_i} = \left\{
  \begin{array}{cc}
    \ket{d_i} & (i \in N_{G}(T))\\
    \prod_{j \in N_{G}(i) \cap N_{G}(T)} Z^{d_j}\ket{+_{\theta_i}} & (i \notin N_{G}(T))\\
  \end{array}\right.
\end{equation}
and $\ket{+_{\theta_j}} = \frac{1}{\sqrt{2}}(\ket{0}+e^{i\theta_j}\ket{1})$.
Bob receives the qubits and applies controlled-$\mathrm{Z}$ gates to the adjoining pairs of qubits in the graph $G$.
(II) For the $i$-th qubit, Alice computes a ciphertext $\delta_{i}^{pk_{i-1}}$ of the $i$-th measurement angle $\delta_i$ by the ${i-1}$-th public key $pk_{i-1}$ and sends it.
Here, the ciphertext $\delta_{1}^{pk_{0}}$ is the plaintext $\delta_{1}$ itself.
Bob decrypts it, measures the qubit with $\delta_i$, and obtains the measurement result $b_i$.
Then, he generates a pair comprising a public key $pk_{i}$ and a secret key $sk_{i}$ with security parameter $n$, encrypts $b_i$ using $pk_{i}$, and sends back $pk_{i}$ and the ciphertext $b_{i}^{pk_{i}}$.
They repeat the above for all qubits.
(III) Alice announces $T$ and the expected results $\lseq{r_t}{t \in T}$.
Then, Bob checks the associated results $\lseq{b_t}{t \in T}$ and announces all secret keys $\lseq{sk_{i}}{i \in V}$ if $b_t = r_t$ for all $t \in T$.
She decrypts all received messages using the secret keys and accepts the outcome of the computation if all secret keys and all messages from Bob are valid and $b_t = r_t$ for all $t \in T$.
(IV) A third party judges Alice to have obtained the correct outcome if Bob reveals the secret keys, all secret keys and messages from Bob are valid, and $b_t = r_t$ for all $t \in T$.

The major difference between my protocol and the FK protocol is that Alice and Bob encrypt their classical messages.
Intuitively, the changes allow third-party verification for the following reason.
The encryption of Bob's messages renders Alice unable to obtain the outcome until Bob verifies that the traps are untouched and she then receives the secret key.
To obtain the secret key, even evil Alice has to announce the true traps and hence cannot cheat; otherwise, Bob aborts the protocol.
A third party can check whether Alice obtains the correct outcome using the disclosed information about the traps.

I will now prove that my protocol is a verifiable BQC protocol that satisfies public verifiability.
Before doing so, I first discuss the four requirements for an encryption scheme, as promised:
(i) Alice can compute $\delta_{i}^{pk_{i-1}}$ from $\phi_i$, $\theta_i$, $X_i$, $Z_i$, $\lseq{r_j}{j \leq i}$, $\lseq{pk_j}{j < i}$, and $\lseq{b_j^{pk_{j}}}{j < i}$.
(ii) Even if Bob sends ill-formed public keys and/or ciphertexts of messages and Alice is unaware of the illegality, her messages encrypted by the public keys do not reveal any information other than the desired angles.
(iii) Any public key has a unique secret key, and anyone can confirm that a given pair of a possibly invalid public key and a possibly invalid secret key is genuine.
(iv) The scheme is semantically secure~\cite{G04,GM82}, which guarantees that any ciphertext includes only negligible information about its plaintext.
More precisely, for any two distinct plaintexts $x, y$, there is only a negligible difference between the probabilities that Alice guesses given ciphertexts of $x$ and $y$ to be the ciphertext of $x$.
The first three requirements are needed to confirm that my protocol is a verifiable BQC protocol, i.e., it is correct, blind, and verifiable; the last requirement will be used to prove that my protocol is publicly verifiable.
While my protocol is independent of a choice of encryption scheme as long as it satisfies the requirements, the reader may question whether such a good encryption scheme really exists.
Fortunately, it is possible to tailor ElGamal encryption~\cite{E85} to the protocol using inattentive evaluations~\cite{SYY99}.
The scheme satisfies requirements~(i--iii) with no additional assumption.
Moreover, it is semantically secure under the widely-held DDH assumption~\cite{TY98}.
The detail of the encryption scheme can be found in the Appendix.

I will now prove that my protocol is a verifiable BQC protocol.
Recall that the underlying FK protocol is a $\epsilon$-verifiable BQC protocol with some $\epsilon$.
Because of the existence of requirement~(i) and the fact that the FK protocol is correct, it is easy to see that my protocol is also correct.
Note that Alice should not be able to compute $\delta_{i}$ itself.
As discussed before, it is essential for my protocol that she cannot obtain the measurement results without the secret keys, but Equation~(\ref{feedforwardeq}) shows she can compute $\lseq{b_i}{}$ from $\lseq{\delta_i}{}$ and values she already has.
Next, I show that my protocol is blind.
The knowledge of Bob in my protocol increases by $\lseq{\delta_i^{pk_{i-1}}}{}$, $T$, and $\lseq{r_t}{}$ from the FK protocol.
The location of traps $T$ and the expected measurement outcomes of traps $\lseq{r_t}{}$ are selected randomly and independently of what Alice computes, so Bob learns nothing about her computation from them.
Requirement~(ii) ensures that what Bob extracts from the messages is, at most, $\lseq{\delta_i}{}$, which he can obtain with the FK protocol.
Hence, his knowledge of Alice's computation is the same as in the FK protocol.
Finally, I show that my protocol is $\epsilon$-verifiable.
In my protocol, Bob receives the messages about the traps.
He can use the information to create fake secret keys so that Alice decrypts $b_{t}^{pk_{t}}$ using the fake secret keys and obtains $r_{t}$, although $b_{t} \neq r_{t}$.
Requirement~(iii) excludes such a possibility.
Under this requirement, Bob cannot use the information about the traps for attacks because Alice already has all the measurement results before she discloses the traps, so Bob cannot change them.
Moreover, any strategy by Bob that uses ill-formed keys or messages fails to deceive Alice.
Because Alice can decrypt all messages at the end of my protocol, it can be assumed without loss of generality that she decides whether she accepts the outcome using only the plaintexts.
This means that Bob can do nothing that he cannot already do in the FK protocol to succeed in forcing Alice to accept the incorrect outcome.
Hence, my protocol is $\epsilon$-verifiable.

I now show that my protocol is $\epsilon$-publicly verifiable.
One of the conditions is easily satisfied.
Because a third party uses the same method as Alice to decide whether to accept the outcome, the third party can detect Bob's deception if and only if Alice can detect it, provided that Alice follows the protocol.
To satisfy the other condition, I need to ensure that Alice cannot obtain any non-negligible partial information without the secret keys.
For this purpose, I weaken Alice.
I assume that Alice has no quantum memory.
The assumption excludes the possibility that she constructs a secret quantum channel and reads out the measurement results directly.
Even if she can send one of the entangled qubits to do that, she cannot maintain the entanglement.
Now, let $\lseq{\delta_{i}}{}$ be the computation.
If Bob follows the protocol and a third party rejects the outcome, Alice cannot obtain the secret keys because a third party always accepts the outcome when Bob announces the secret keys.
Hence, in such a case, all she obtains is the public keys $\lseq{pk_{i}}{}$ and the encrypted measurement results $\lseq{b_{i}^{pk_{i}}}{}$.
Note that the $i$-th public key is chosen after the $i$-th measurement $b_{i}$, so $pk_{i}$ is independent of $b_{i}$ and any other information about $\lseq{\delta_{i}}{}$.
Let us consider a case where Bob sends the ciphertexts of $1$ by $\lseq{pk_{i}}{}$ instead of $\lseq{b_{i}^{pk_{i}}}{}$.
If Alice can extract some non-negligible information about $\lseq{\delta_{i}}{}$ from $\lseq{b_{i}^{pk_{i}}}{}$, she can distinguish between that case and a case where Bob actually sends $\lseq{b_{i}^{pk_{i}}}{}$ with a non-negligible probability.
However, this contradicts requirement~(iv).
Therefore, I conclude that the protocol is $\epsilon$-publicly verifiable.

In this paper, I proposed a new verifiable BQC protocol based on the FK protocol.
In comparison, my protocol preserves all properties of the FK protocol without any additional assumptions; it also offers public verifiability.
However, to achieve public verifiability, my protocol requires messages to be much longer.
The disadvantage is not serious; although the messages are long in my protocol, their total size is polynomial in the size of the messages in the FK protocol.
This means that any computation efficiently executable in the FK protocol is also efficiently executable in my protocol.
Furthermore, homomorphic encryption schemes~\cite{GM82,G09,OPP14} can be used and the size can be reduced if blindness is made to hold under some assumptions~\cite{GOS06} or we give up perfect blindness~\cite{DKL12}.

\acknowledgments
The author thanks Takahiro Kubota, Tomoyuki Morimae, Yoshihiko Kakutani, Joseph Fitzsimons, and Fran\c{c}ois Le Gall for insightful discussions and comments.
This work was supported by JSPS Grant-in-Aid for JSPS Fellows Grant No.\ $26\mathrel{\cdot}9148$.

\appendix

\section{Definition of public verifiability}
Let $\xi$ be a real number such that $0 \leq \xi< 1$. 
A BQC protocol is said to be $\xi$-publicly verifiable if there is a polynomial-time algorithm, which is called the public verification procedure, that computes ``accept'' or ``reject'' from all classical messages in the protocol and satisfies the following two conditions when the computational ability of Alice does not exceed one of a probabilistic classical computer.
\begin{enumerate}
\item The probability that the procedure outputs ``accept'' but Alice does not obtain the correct outcome is not more than $\xi$ provided that she follows the protocol.
\item
  There exists a simulator of Alice such that for any her computation $\lseq{\delta_{i}}{}$ and any function $h_{n}, f_{n}$, there is only a negligible difference between the probability that Alice correctly computes $f_{n}(\lseq{\delta_{i}}{})$ from $h_{n}(\lseq{\delta_{i}}{})$, $\lseq{\delta_{i}}{}$, and the messages she obtains in an execution of the protocol with honest Bob where the public verification procedure outputs ``reject'' and the probability that the simulator correctly computes $f_{n}(\lseq{\delta_{i}}{})$ from $h_{n}(\lseq{\delta_{i}}{})$ and $\lseq{\delta_{i}}{}$.
  Here, the difference is negligible with respect to $n$, and all $|\lseq{\delta_{i}}{}|$, $|h_{n}(\lseq{\delta_{i}}{})|$, and $|f_{n}(\lseq{\delta_{i}}{})|$ are polynomially bounded with respect to $n$.
\end{enumerate}

It is easy to modify the proof in the main body to prove that my protocol is publicly verifiable with this definition.

\section{Encryption scheme}
ElGamal encryption is one of the best-known forms of probabilistic, public-key encryption scheme, and uses the difficulty in computing discrete logarithms.
Although a quantum computer computes discrete logarithms efficiently~\cite{S97}, Alice cannot use a quantum computer to decrypt ciphertexts as discussed above.
ElGamal encryption satisfies all requirements except the first one~\cite{TY98}.
While ElGamal encryption is known to be multiplicatively homomorphic, it is ill-defined in my setting.
As I use bitwise encoding, it is enough for the computation~(\ref{feedforwardeq}) to evaluate a formula $(\alpha_{1} \vee \alpha_{2}) \oplus (\alpha_{3} \vee \alpha_{4})$, where $\alpha_{j} \in \{0,1, b_{X_i}, b_{Z_i}, \neg b_{X_i}, \neg b_{Z_i}\}$.
To evaluate it, I used inattentive evaluations, which render secret evaluations of log-depth circuits possible using inductive construction.
Inattentive evaluations require Bob to send $(0, 1)$ or $(1, 0)$ instead of $0$ or $1$, respectively, so evil Bob sends ill-formed messages such as $(0, 0)$ and possibly obtains partial information about the dependency of the above logical formula on $b_{X_i}$ and $b_{Z_i}$.
Although a non-interactive zero-knowledge proof~\cite{BDMP91} was employed in the original paper~\cite{SYY99}, I do not use it here, because it requires the additional assumption that all parties share a common reference string.
I modified the method such that Alice encodes a bit using the received messages and uses it instead of the plain bit.
This can make her message completely meaningless when Bob sends an ill-formed message.

Specifically, with a prime $p_i$ where $2p_i+1$ is also a prime, the $i$-th public key $pk_{i}$ is a trio of the cyclic subgroup $H_i$ of order $p_i$ of ${\Z}_{2p_i+1}$, a generator $g_i$ of $H_i$, and a randomly selected element $g_i^{x_i}$ of $H_i$.
The associated secret key $sk_{i}$ is $x_i$.
Let $0_{i}^{*} \equiv (g_i^{{x_i}r}, g_i^r)$ and $1_{i}^{*} \equiv (g_i^{m}g_i^{{x_i}r}, g_i^{r})$ with randomly selected $r$ and $m \neq 0$.
The ciphertext $b_{i}^{pk_{i}}$ of the $i$-th measurement result $b_{i}$ is $(b_i^{*}, c_i^{*})$ where $b = b_{i}$ and $c = 1 - b_{i}$.
Next, Alice computes $\delta_{i}^{pk_{i-1}}$.
As discussed above, she inductively computes $(\alpha_{1} \vee \alpha_{2}) \oplus (\alpha_{3} \vee \alpha_{4})$.
She encodes a bit value using given ciphertexts at the zeroth level, computes bit summation of them at the first level, computes logical disjunction of them at the second level, and finally takes bit summation of them at the third level.
At the zero level, Alice encodes a bit into four pairs of bits: she encodes $0$ into four pairs three of which are $(0, 0)$ or $(1, 1)$ and one of which is $(0, 1)$ or $(1, 0)$; $1$ into three $(0, 1)$ or $(1, 0)$ and one $(0, 0)$ or $(1, 1)$.
Specifically, she creates $\mathbf{0}_{i}$, $\mathbf{1}_{i}$, $\mathbf{b}_{i}$, and $\mathbf{\neg{b}}_i$ where
\begin{align}
  \mathbf{0}_{i} & \equiv \{(b_{i}^{*}, 0_{i}^{*}), (c_{i}^{*}, 0_{i}^{*}), (b_{i}^{*}, b_{i}^{*}), (b_{i}^{*}, b_{i}^{*})\}\\
  \mathbf{1}_{i} & \equiv \{(b_{i}^{*}, 0_{i}^{*}), (c_{i}^{*}, 0_{i}^{*}), (b_{i}^{*}, c_{i}^{*}), (b_{i}^{*}, c_{i}^{*})\}\\
  \mathbf{b}_{i} & \equiv \{(b_{i}^{*}, 0_{i}^{*}), (b_{i}^{*}, 0_{i}^{*}), (b_{i}^{*}, b_{i}^{*}), (b_{i}^{*}, c_{i}^{*})\}\\
  \mathbf{\neg{b}}_{i} & \equiv \{(c_{i}^{*}, 0_{i}^{*}), (c_{i}^{*}, 0_{i}^{*}), (b_{i}^{*}, b_{i}^{*}), (b_{i}^{*}, c_{i}^{*})\}.
\end{align}
In the first level, Alice uses an $i$-length bit sequence whose bit summation denotes its value.
For $b_{X_i}$, she creates an $i$-length sequence, $\lseq{a_j}{j < i}$, where $a_j$ is $\mathbf{b}_j$ if $j \in X_i$; otherwise, it is $\mathbf{0}_j$.
To denote the negation of it, she just flips the value of $a_0$.
The encoding of the second level is the same flavor as that in the zero level.
She encodes $\alpha \vee \beta$ into $\{(\alpha, 0), (\beta, 0), (\alpha, \beta), (1, 0)\}$.
Finally, in the third level, she forms a pair as she did in the first level, and she finishes evaluating the desired formula.
Before sending it, Alice randomize it so that Bob cannot obtain any information other than $\delta_{i}$.
For the given ciphertext $(g_j^{s}, g_j^{r})$, she can create a new random ciphertext $(g_j^{sy+xz}, g_j^{ry+z})$ of the same plaintext without knowing the plaintext with randomly selected $y \neq 0$ and $z$.
She permutes pairs in the zeroth and second levels, flips even number of bits in the first level, and, in the second level, she both permutes pairs and flips even number of bits.

It is easy to check that Bob can extract $\delta_{i}$ from the above $\delta_{i}^{pk_{i-1}}$.
The validity of public key can be checked in polynomial-time and the secret key is unique.
Therefore, the above construction satisfies the third requirement.
Next, I will see the construction satisfies the second requirement.
If Bob sends the well-formed messages, the property of inattentive evaluation guarantees that Bob obtain nothing other than $\delta_{i}$~\cite{SYY99}.
Note that he knows that Alice computes $(\alpha_{1} \vee \alpha_{2}) \oplus (\alpha_{3} \vee \alpha_{4})$ and the shape of the formula is not secret.
The validity of a ciphertext of ElGamal encryption is easily checked.
Therefore, all evil Bob can do is sending $(0_i^{*}, 0_i^{*})$ or $(1_i^{*}, 1_i^{*})$.
However, in the case, $\mathbf{0}_{j} = \mathbf{1}_{j} = \mathbf{b}_{j} = \mathbf{\neg{b}}_j$ and he gains nothing.
The construction is semantically secure because ElGamal encryption is semantically secure under decisional Diffie-Hellman (DDH) assumption~\cite{TY98}.
Therefore, the above construction satisfies all requirements.

\bibliographystyle{apsrev}
\bibliography{Paper}

\end{document}